\documentclass[preprintnumbers,amsmath,11pt,amssymb,floatfix,superscriptaddress,nofootinbib]{revtex4}
\usepackage{graphicx}
\usepackage{epsfig}
\usepackage{bm}
\usepackage{amsfonts}

\input epsf

\begin{document}

\title{\Large The generalized second law of thermodynamics in $f(\cal R)$ gravity for various choices of scale factor}

\author{\bf Rahul Ghosh$^1$\footnote{ghoshrahul3@gmail.com}, Surajit
Chattopadhyay$^2$\footnote{surajit$_{_{-}}2008$@yahoo.co.in}}

\affiliation{$^1$ Department of Mathematics, Bhairab Ganguly
College, Kolkata-700 056, India.\\
$^2$Department of Computer Application (Mathematics Section),
Pailan College of Management and Technology, Bengal Pailan Park,
Kolkata-700 104, India.}

\date{\today}

\begin{abstract}
The present study is aimed at investigating the validity of the
generalized second law (GSL) of thermodynamics in $f(\cal R)$
gravity. Choosing $f(\cal R)=R+\xi R^{\mu}+\zeta R^{-\nu}$
[following Phys Rev D, \textbf{68} 123512 (2003)] we have computed
the time derivatives of total entropy for various choices of scale
factor pertaining to emergent, intermediate and logamediate
scenarios of the universe. We have taken into account the radii of
hubble, apparent, particle and event horizons while computing the
time derivatives of entropy under various situations being
considered. After analyzing through the plots of time derivative
of total entropy against cosmic time it is observed that the
derivative always stays at positive level that indicate the
validity of GSL of thermodynamics in the $f(\cal R)$ gravity
irrespective of the choices of scale factor and enveloping
horizon.
\end{abstract}


\maketitle

\section{\normalsize\bf{Introduction}}

Accelerated expansion of the universe is well documented in
literature (detailed discussion is available in \cite{sami} and
references therein). Approximately $76\%$ of the energy content of
the Universe is not dark or luminous matter but it is instead a
mysterious form of dark energy that is exotic, invisible, and
unclustered \cite{Riess}. In order to explain the origin of this
form of matter three main classes of models for this acceleration
exist \cite{Kluson}: (1) A cosmological constant $\Lambda$ (2)
Dark Energy and (3) Modified Gravity. The last class of models
known as extended theories of gravity corresponds to the
modification of the action of the gravitational fields
\cite{Kluson}. These theories are based on the idea of an
extension of the Einstein Hilbert action by adding higher order
curvature invariants. Modified gravity theories have been reviewed
in \cite{Nojirireview, Mukhoyama, Paul}. Nojiri and Odintsov
\cite{Nojiri1} suggested $f(\cal R)$ gravity characterized by the
presence of effective cosmological constant epochs in such a way
that early time inflation and late time cosmic acceleration are
mutually unified within a single model. In another work, Nojiri
and Odintsov \cite{Nojiri2} proposed another class of modified
$f(\cal R)$ to unify ${\cal R}^m$ inflation with $\Lambda$CDM era.
Chattopadhyay and Debnath \cite{chattopadhyay1} considered $f(\cal
R)$ gravity in an universe characterized by a special form scale
factor known as emergent scenario and observe that and concluded
that the EoS parameter behaves like quintessence in this
situation.

In the present work are going to investigate the GSL of
thermodynamics for various choices of the scale factor $a$. The
choices are named as ``emergent",
 ``intermediate" and ``logamediate". The physical aspects behind such choices are well documented in the literature
 \cite{Mukherjee,barrow1,barrow2}. However, for the sake of convenience we shall give a brief overview in a subsequent section. In this
  work the thermodynamic consequences of the universe for the said choices of scale factor would be examined in a modified gravity theory named as $f(\cal R)$ gravity that
has gained immense interest in recent times. In the present work
we have extended the study of the reference \cite{chattopadhyay1}
by to the investigation of the generalized second law (GSL) of
thermodynamics in emergent scenario with the universes enveloped
by Hubble, apparent, Particle and Event horizons respectively. In
the remaining part of the paper the radii of the said horizons are
denoted by $R_{H}$, $R_{A}$, $R_{P}$ and $R_{E}$ respectively.
Validity of the GSL implies that the sum of the time derivatives
of the internal entropy and entropy on the horizon is
non-negative. Hence, the primary objective of this work is to
discern whether
$\dot{S}_{total}=\dot{S}_{internal}+\dot{S}_{horizon}\geq 0$ holds
for the situations under consideration. The GSL would be
investigated based on the first law of thermodynamics. Relevance
of the laws of thermodynamics in cosmology was discussed by
\cite{Lima, Silva}. Validity of GSL in various DE candidates and
their interactions have been discussed in several papers like
\cite{Sheykhi,chattopadhyay2,Setare2,Setare3,Setare4,Elizalde,Izquierdo}.
The works on the validity of the GSL in modified gravity theories
include \cite{jamil2,Akbar2,Akbar3,Cai2,Wu}. In the reference
\cite{bambageng} studied thermodynamics of the apparent horizon in
$f(\cal R)$ gravity and it was demonstrated that an $f(\cal R)$
gravity can realize a crossing of the phantom divide and can
satisfy the second law of thermodynamics in the effective phantom
phase as well as non-phantom one. In another work, \cite{Akbar3}
studied the thermodynamic behavior of field equations for $f(\cal
R)$ gravity. In the present work, we have taken different choices
for the scale factor and examined whether the GSL holds for those
choices. Details are presented in the subsequent sections.
\\\\

\section{The generalized second law}
In this section we are going to examine whether the generalized second law (GSL)will hold for various choices of scale factor and on various horizons under
$f(\cal R)$ gravity. The basic necessity for the validity of GSL is that the time derivative of the total entropy $\dot{S}_{Total}=\dot{S}_{H}+\dot{S}\geq
0$, where $\dot{S}$ indicates the time derivative of normal entropy and
$\dot{S}_{H}$ indicates the horizon entropy \cite{Izquierdo}.\\
The first law of thermodynamics (Clausius relation) on the horizon
is defined as $T_{X}dS_{X}=\delta Q=-dE_{X}$. From the unified
first law, we may obtain the first law of thermodynamics as

\begin{equation}
T_{X}dS_{X}=4\pi R_{X}^{3}H (\rho+p)dt
\end{equation}

where, $T_{X}$ and $R_{X}$ are the temperature and radius of the
horizons under consideration in the equilibrium thermodynamics.
Subsequently, the time derivative of the entropy on the horizon
can be derived as

\begin{equation}
\dot{S}_{X}=\frac{4\pi R_{X}^{3}H}{T_{X}}(\rho+p)
\end{equation}

Finally, we can get the time derivative of total entropy as
\cite{arundhati}

\begin{equation}
\dot{S}_{Total}=\dot{S}_{X}+\dot{S}_{IX}=\frac{ R_{X}^{2}}{GT_{X}}\left(\frac{k}{a^{2}}-\dot{H}\right)\dot{R}_{X}
\end{equation}

Our target is to investigate whether $\dot{S}_{X}+\dot{S}_{IX}\geq
0$ holds.
\\\\
\subsection{Basic equations of $f(\cal R)$ gravity}
The action of $f(\cal R)$ gravity is given by \cite{bambageng}

\begin{equation}
S=\int d^{4}x\sqrt{-g}\left[\frac{f(\cal R)}{2\kappa^{2}}+\mathcal{L}_{matter}\right]
\end{equation}
where $g$ is the determinant of the metric tensor $g_{\mu\nu}$, $\mathcal{L}_{matter}$ is the matter Lagrangian and $\kappa^{2}=8\pi G$. The $f(\cal R)$ is
a non-linear function of the Ricci curvature $cal R$ that incorporates corrections to the Einstein-Hilbert action which is instead described by a linear
function $f(\cal R)$. The gravitational field equations in this theory are \cite{bambageng}

\begin{equation}
H^{2}+\frac{k}{a^{2}}=\frac{\kappa^{2}}{3f'(\cal R)}(\rho+\rho_{c})
\end{equation}

\begin{equation}
\dot{H}-\frac{k}{a^{2}}=-\frac{\kappa^{2}}{2f'(\cal R)}(\rho+p+\rho_{c}+p_{c})
\end{equation}

where $\rho_{c}$ and $p_{c}$ can be regarded as the energy density and pressure generated due to the difference of $f(\cal R)$ gravity from general
relativity given by \cite{bambageng}

\begin{equation}
\rho_{c}=\frac{1}{8 \pi f'}\left[- \frac{f - R f'}{2} -3 H f''\dot{R}\right]
\end{equation}

\begin{equation}
p_{c}=\frac{1}{8 \pi f'}\left[\frac{f -  {\cal R} f'}{2} + f'' \ddot{\cal R} + f''' \ddot{\cal R}^{2} + 6 f'' \dot{\cal R}\right]
\end{equation}

where, the scalar tensor $ {\cal R}=-6\left(\dot{H} + 2 H^{2} + \frac{k}{a^{2}}\right)$.
\\\\
\subsection{The choices of scale factor $a(t)$}
In this paper we have considered three forms of the scale factor $a(t)$ in the $f(\cal R)$ gravity to investigate the validity of the GSL of
thermodynamics. The three choices, in literature, dubbed as ``emergent", ``intermediate" and ``logamediate" respectively, are given by
\begin{enumerate}
    \item \textbf{Emergent}: $a(t)=A\left(\eta+e^{B t}\right)^{n}$
    with $A>0,~~B>0,~~\eta>0,~~n>1$ \cite{Mukherjee}.
    \item \textbf{Intermediate}: $a(t)=exp (B t^{\beta})$ with
    $B>0;~~0<\beta<1$ \cite{barrow1}.
    \item  \textbf{Logamediate}: $a(t)=exp(A(\ln
    t)^{\alpha})$ with $A\alpha>0,~~\alpha>1$ \cite{barrow2}.\\
\end{enumerate}
For the above choices of scale factor the forms of the Hubble
parameter $H$ are the following

\begin{equation}
H=\frac{Bn e^{Bt}}{\eta+e^{B t}};~~~~~H=B\beta
t^{-1+\beta};~~~~~H=\frac{A\alpha (\ln t)^{-1+\alpha}}{t}
\end{equation}

It is clear from the above that we are first choosing various forms of scale factor and subsequently investigating the GSL in the corresponding scenarios.
This `reverse' way of investigations had earlier been used extensively by Ellis and Madsen \cite{ellis}, who chose various forms of scale factor and then
found out the other variables from
the field equations. \\
We choose the function $f(\cal R)$ as \cite{Nojiri3}
\begin{equation}
f(\cal R)=\cal R+\xi \cal R^{\mu}+\zeta \cal R^{-\nu}
\end{equation}

We obtain the Ricci scalar $\cal R$ for the above three choices of scale factor leading to the forms of $H$ obtained in (11). Subsequently we obtain
$f(\cal R)$ for all of the above choices as functions of time $t$. The radii of the various enveloping horizons of the universe are given below.

The radius of the apparent horizon is given by
\begin{equation}
R_{A}=\frac{1}{\sqrt{H^{2}+\frac{k}{a^{2}}}}\nonumber
\end{equation}
If we use $k=0$, then we get the radius of the Hubble horizon $R_{H}=\frac{1}{H}$. The radii of the particle $R_{P}$ and the event $R_{E}$ horizons are
given by
\begin{equation}
R_{P}=a\int_{0}^{t}\frac{dt}{a}~~~~ \text{and}~~~~~
R_{E}=a\int_{0}^{\infty}\frac{dt}{a}\nonumber
\end{equation}

Discussions on the above radii of different horizons are available in \cite{arundhati}.

Using the the above forms of scale factors the Ricci scalar $\cal R$ is reconstructed as follows:

For ``emergent" scenario:
\begin{equation}
\begin{array}{c}
{\cal
R}=6\left[-\frac{B^2e^{2Bt}n}{(e^{Bt}+\eta)^2}+\frac{2B^2e^{2Bt}n^2}{(e^{Bt}+\eta)^2}+\frac{B^2e^{Bt}n}{e^{Bt}+\eta}+\frac{k(e^{Bt}+\eta)^{-2n}}{A^2}\right]
\end{array}
\end{equation}

For ``intermediate" scenario:
\begin{equation}
\begin{array}{c}
{\cal R}=6\left[e^{-2Bt^\beta}k+Bt^{-2+\beta}(-1+\beta)\beta+2B^2t^{-2+2\beta}\beta^2\right]
\end{array}
\end{equation}

For ``logamediate" scenario:
\begin{equation}
\begin{array}{c}
{\cal R}=6\left[e^{-2A(\ln t)^\alpha}k+\frac{A(-1+\alpha)\alpha(\ln t)^{-2+\alpha}}{t^2}-\frac{A\alpha(\ln t)^{-1+\alpha}}{t^2}+\frac{2A^2\alpha^2(\ln
t)^{-2+2\alpha}}{t^2}\right]
\end{array}
\end{equation}

 Now we have discussed the validity of the GSL of
thermodynamics with the various choices of scale factor by obtaining the time derivatives of total entropy from for the universe enveloped by the different
horizons and then plotted the time derivatives of the total entropy against cosmic time $t$ to get the following twelve graphs shown in the figure 1 to 12
[figure 1 to 3 ( for hubble horizon), figure 4 to 6 ( for apparent horizon), figure 7 to 9 ( for particle horizon) and figure 10 to 12 ( for event
horizon)]. In all the plots we find that $\dot{S}_{total}$ is staying in the positive level. This indicates the validity of GSL of thermodynamics in all
scenarios of the universe enveloped by the hubble, apparent, particle and event horizons.
\\

\begin{figure}
\includegraphics[height=2.8in]{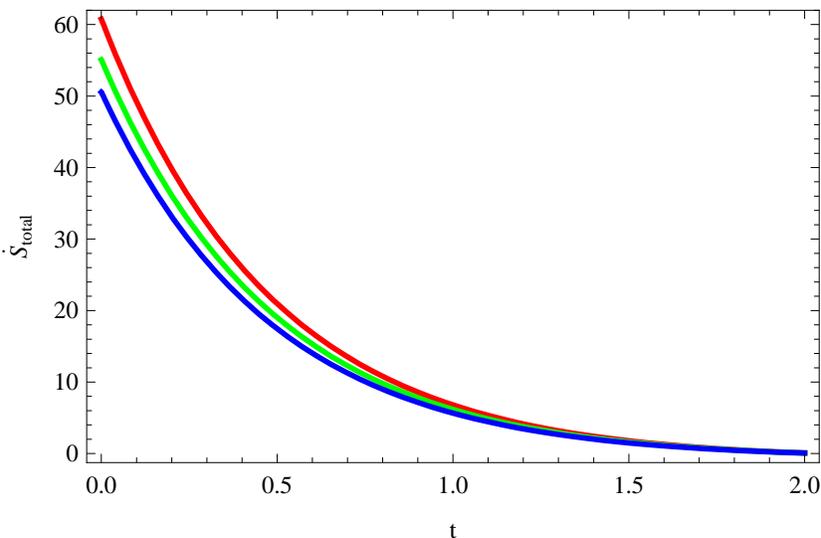}\\
\caption{ plots the time derivative of total entropy $\dot{S}_{total}$ against cosmic time $t$ for the universe enveloped by \emph{Hubble horizon} in the
$f(\cal R)$ gravity in the \emph{``Emergent scenario"}. The red, green and blue lines correspond to $k=-1,~+1\text{and}~k=0$ respectively.}
\end{figure}
\begin{figure}
\includegraphics[height=2.8in]{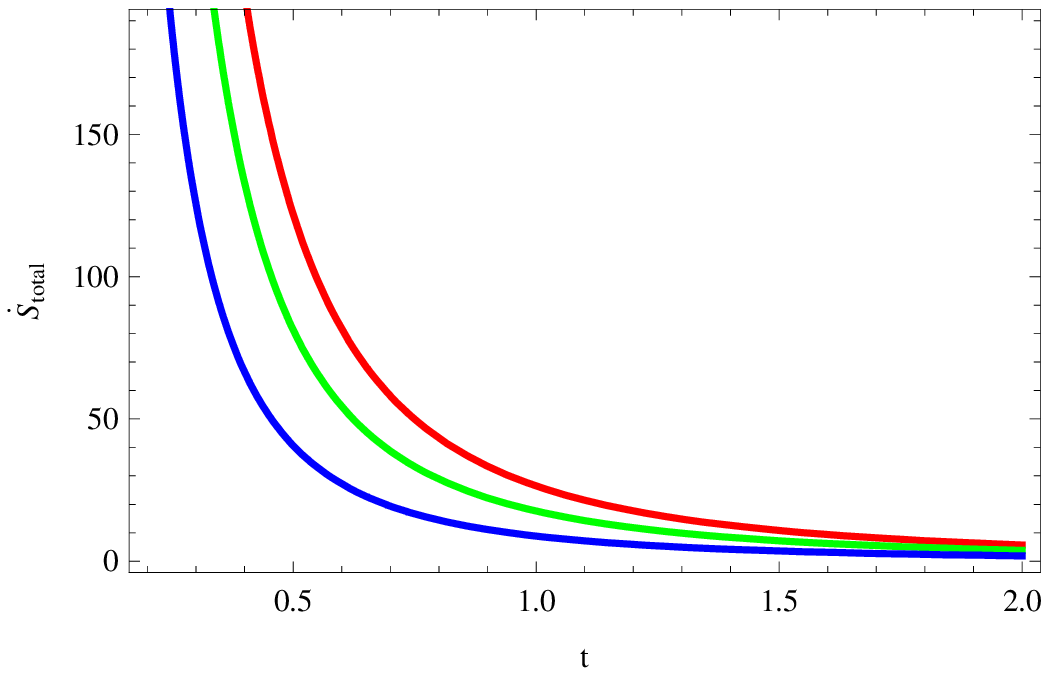}\\
\caption{plots the time derivative of total entropy $\dot{S}_{total}$ against cosmic time $t$ for the universe enveloped by\emph{ Hubble horizon} in the
$f(\cal R)$ gravity in the \emph{``Intermediate scenario"}. The red, green and blue lines correspond to $k=-1,~+1\text{and}~k=0$ respectively.}
\end{figure}
\begin{figure}
\includegraphics[height=2.8in]{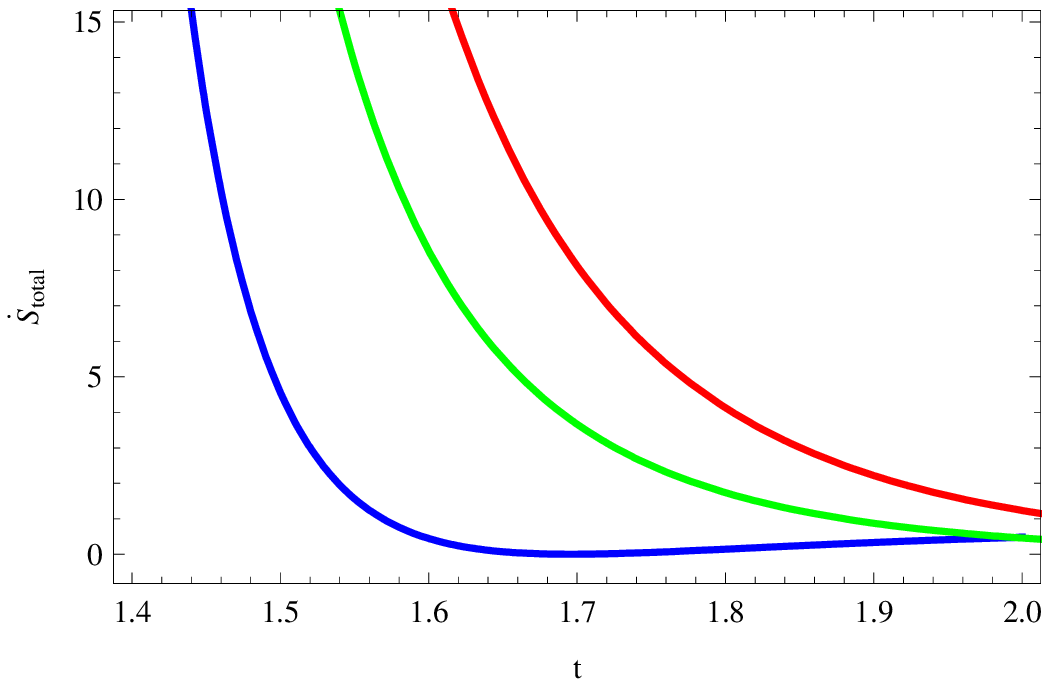}\\
\caption{ plots the time derivative of total entropy $\dot{S}_{total}$ against cosmic time $t$ for the universe enveloped by \emph{Hubble horizon} in the
$f(\cal R)$ gravity in the \emph{``Logamediate scenario"}. The red, green and blue lines correspond to $k=-1,~+1\text{and}~k=0$ respectively.}
\end{figure}
\begin{figure}
\includegraphics[height=2.0in]{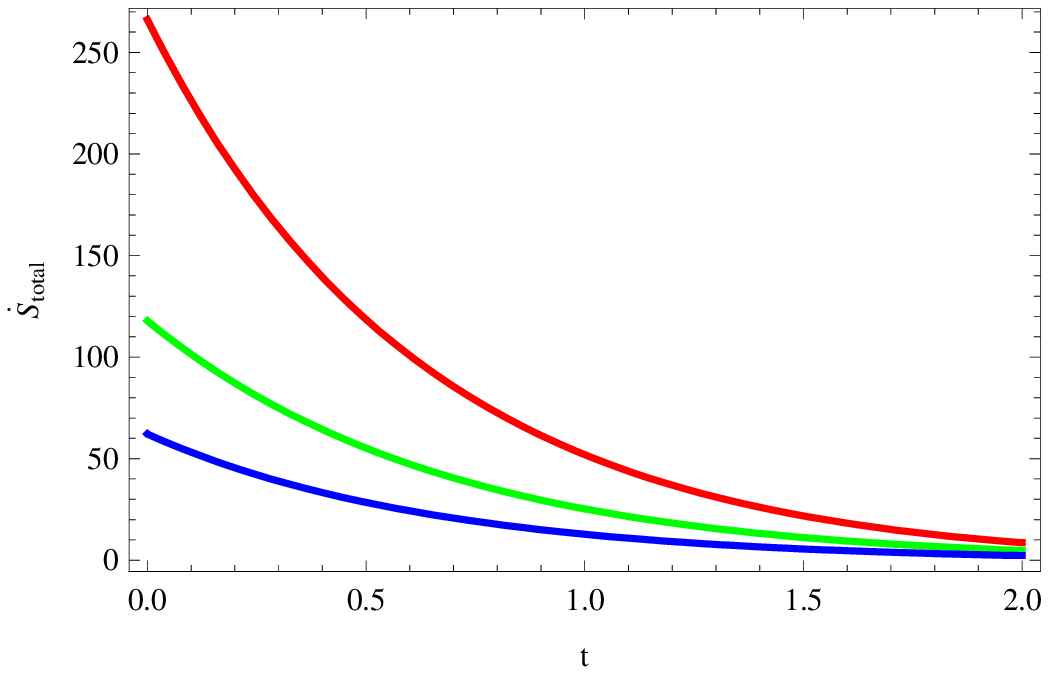}\\
\caption{ plots the time derivative of total entropy $\dot{S}_{total}$ against cosmic time $t$ for the universe enveloped by \emph{Apparent horizon} in the
$f(\cal R)$ gravity in the \emph{``Emergent scenario"}. The red, green and blue lines correspond to $k=-1,~+1\text{and}~k=0$ respectively.}
\end{figure}
\begin{figure}
\includegraphics[height=2.0in]{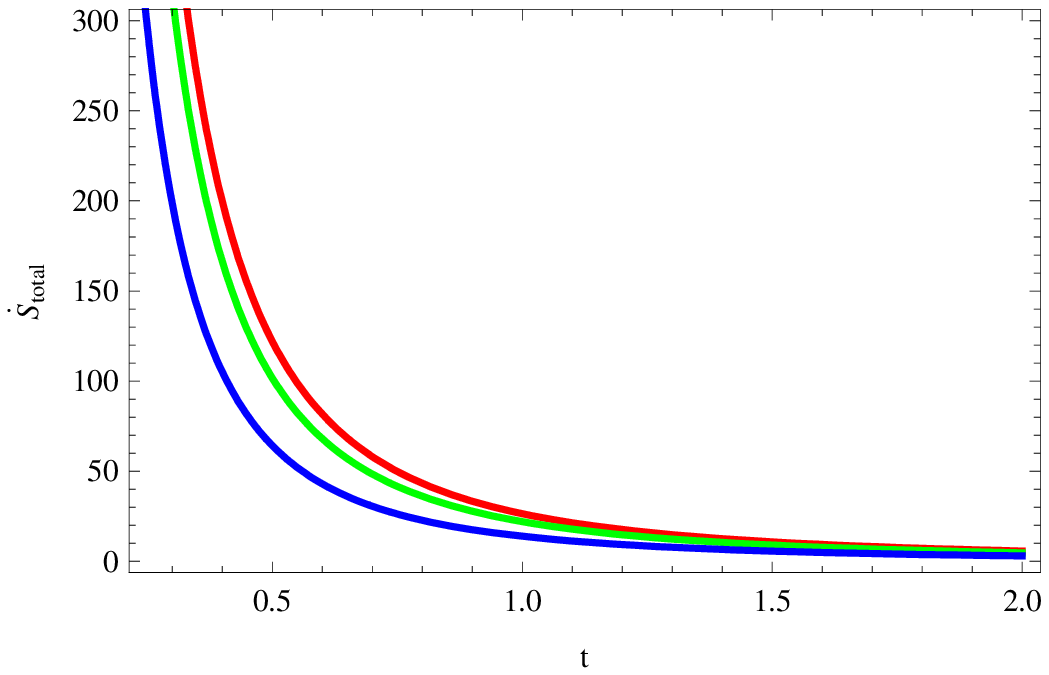}\\
\caption{ plots the time derivative of total entropy $\dot{S}_{total}$ against cosmic time $t$ for the universe enveloped by \emph{Apparent horizon} in the
$f(\cal R)$ gravity in the \emph{``Intermediate scenario"}. The red, green and blue lines correspond to $k=-1,~+1\text{and}~k=0$ respectively.}
\end{figure}
\begin{figure}
\includegraphics[height=2.0in]{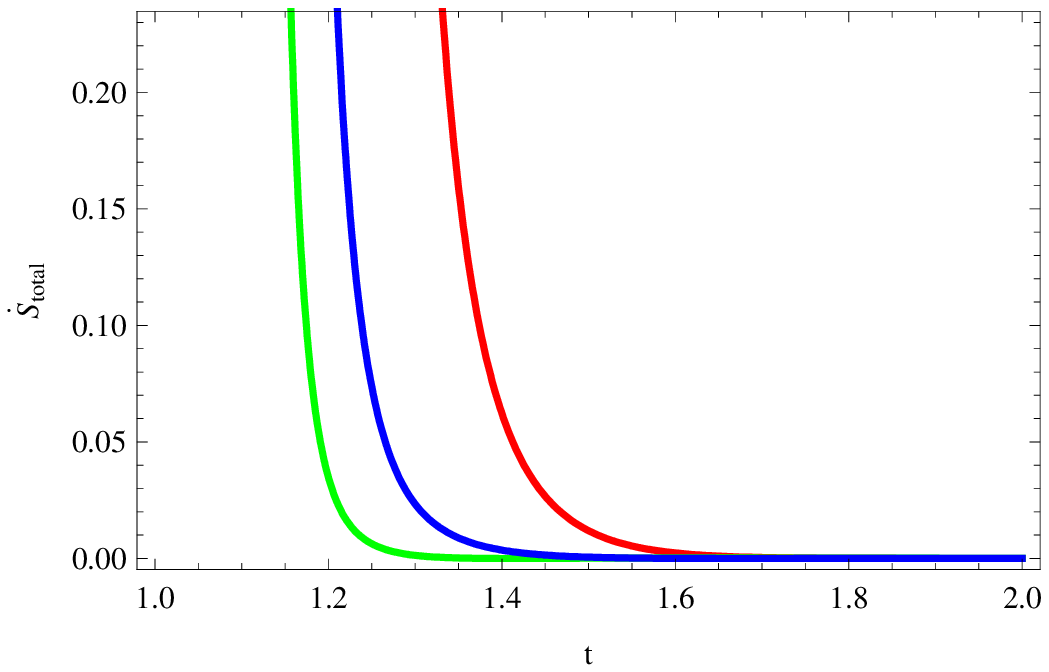}\\
\caption{ plots the time derivative of total entropy $\dot{S}_{total}$ against cosmic time $t$ for the universe enveloped by \emph{Apparent horizon} in the
$f(\cal R)$ gravity in the \emph{``Logamediate scenario"}. The red, green and blue lines correspond to $k=-1,~+1\text{and}~k=0$ respectively.}
\end{figure}
\begin{figure}
\includegraphics[height=2.0in]{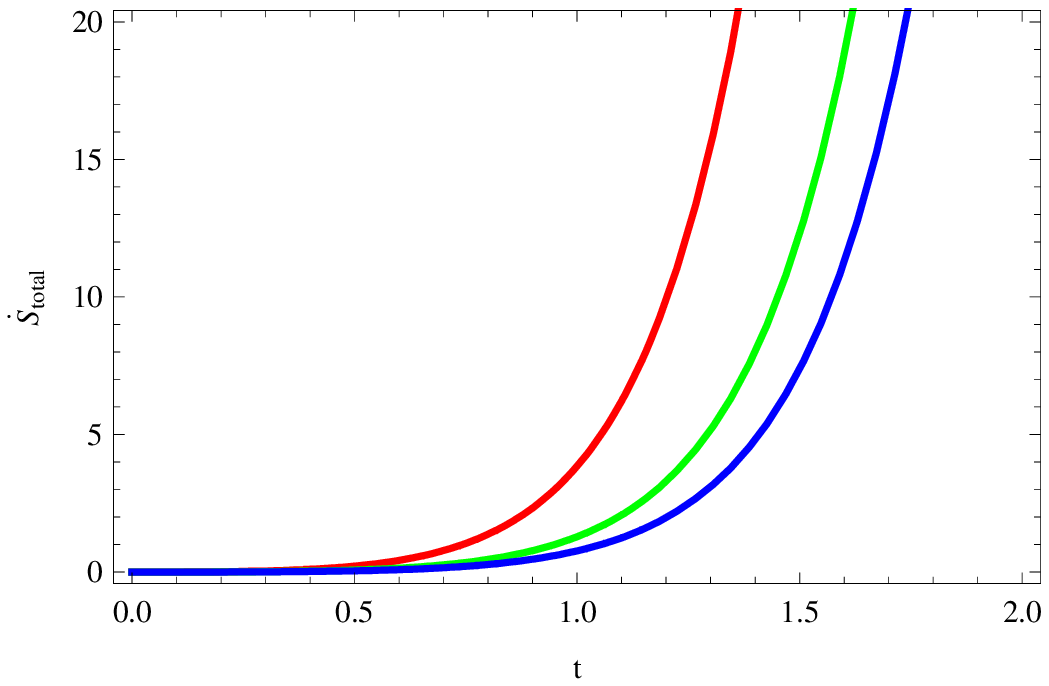}\\
\caption{ plots the time derivative of total entropy $\dot{S}_{total}$ against cosmic time $t$ for the universe enveloped by \emph{Particle horizon }in the
$f(\cal R)$ gravity in the \emph{``Emergent scenario"}. The red, green and blue lines correspond to $k=-1,~+1\text{and}~k=0$ respectively.}
\end{figure}
\begin{figure}
\includegraphics[height=2.0in]{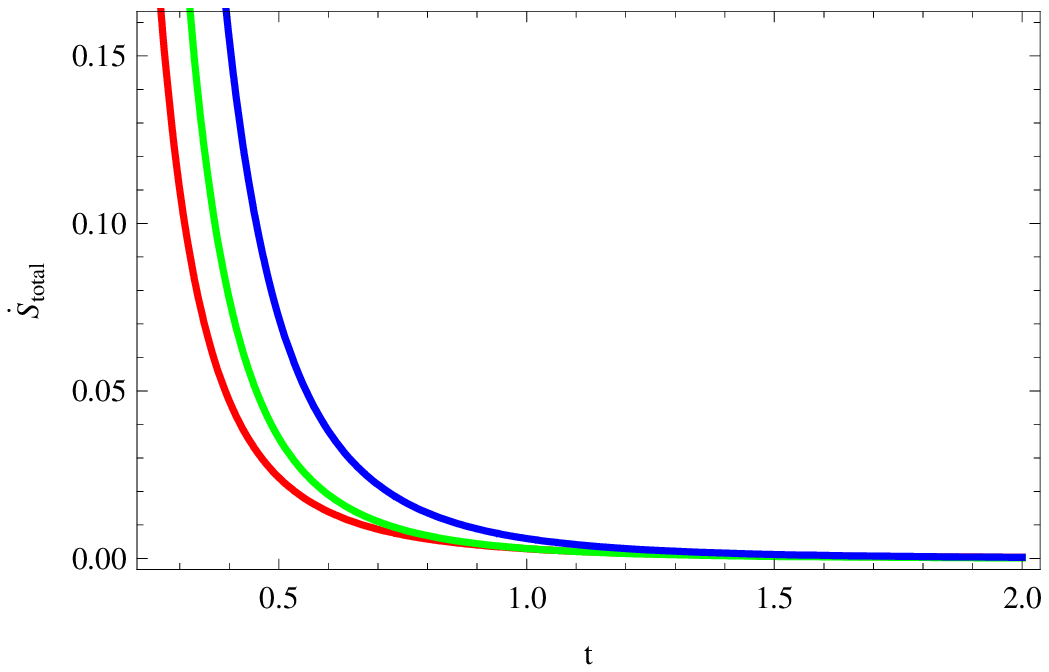}\\
\caption{ plots the time derivative of total entropy $\dot{S}_{total}$ against cosmic time $t$ for the universe enveloped by \emph{Particle horizon} in the
$f(\cal R)$ gravity in the \emph{``Intermediate scenario"}. The red, green and blue lines correspond to $k=-1,~+1\text{and}~k=0$ respectively.}
\end{figure}
\begin{figure}
\includegraphics[height=2.0in]{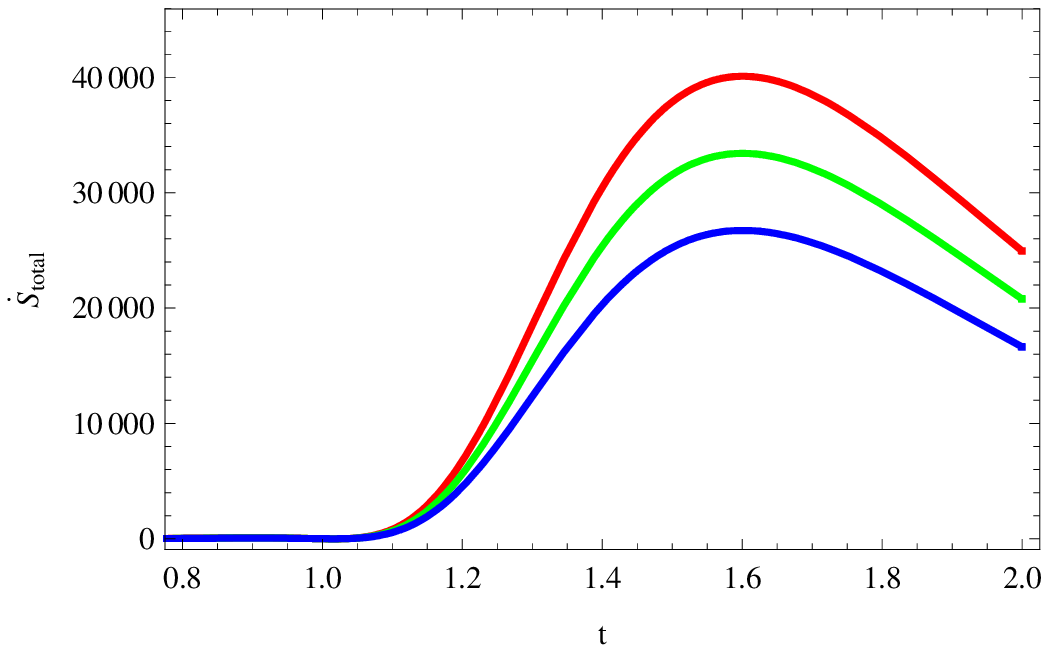}\\
\caption{ plots the time derivative of total entropy $\dot{S}_{total}$ against cosmic time $t$ for the universe enveloped by \emph{Particle horizon} in the
$f(\cal R)$ gravity in the \emph{``Logamediate scenario"}. The red, green and blue lines correspond to $k=-1,~+1\text{and}~k=0$ respectively.}
\end{figure}
\begin{figure}
\includegraphics[height=2.0in]{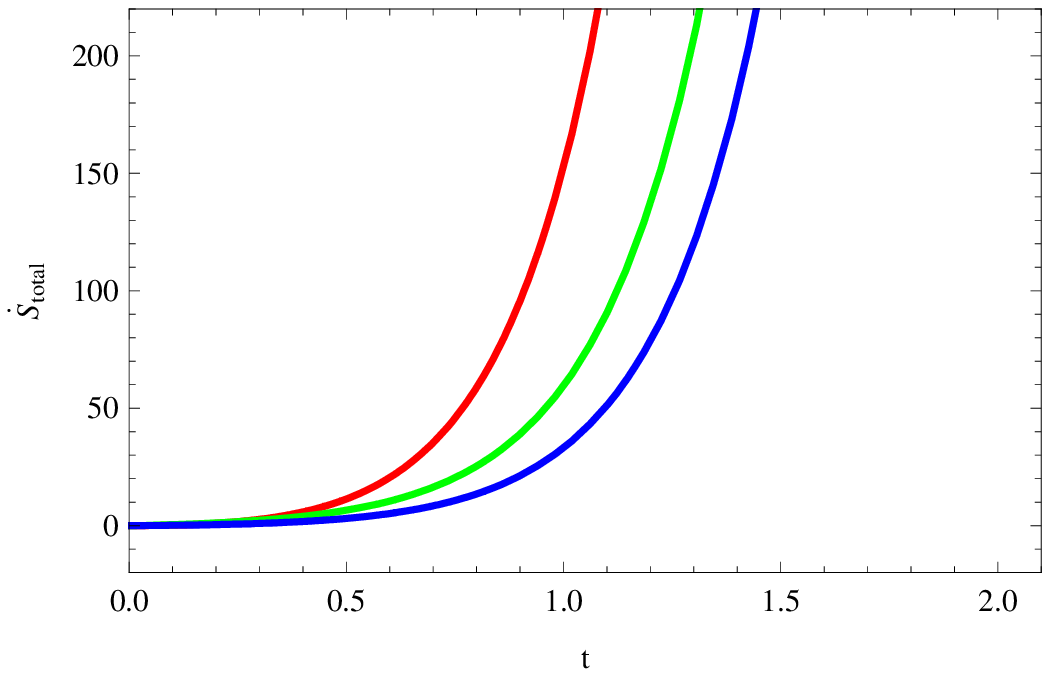}\\
\caption{ plots the time derivative of total entropy $\dot{S}_{total}$ against cosmic time $t$ for the universe enveloped by \emph{Event horizon }in the
$f(\cal R)$ gravity in the \emph{``Emergent scenario"}. The red, green and blue lines correspond to $k=-1,~+1\text{and}~k=0$ respectively.}
\end{figure}
\begin{figure}
\includegraphics[height=2.0in]{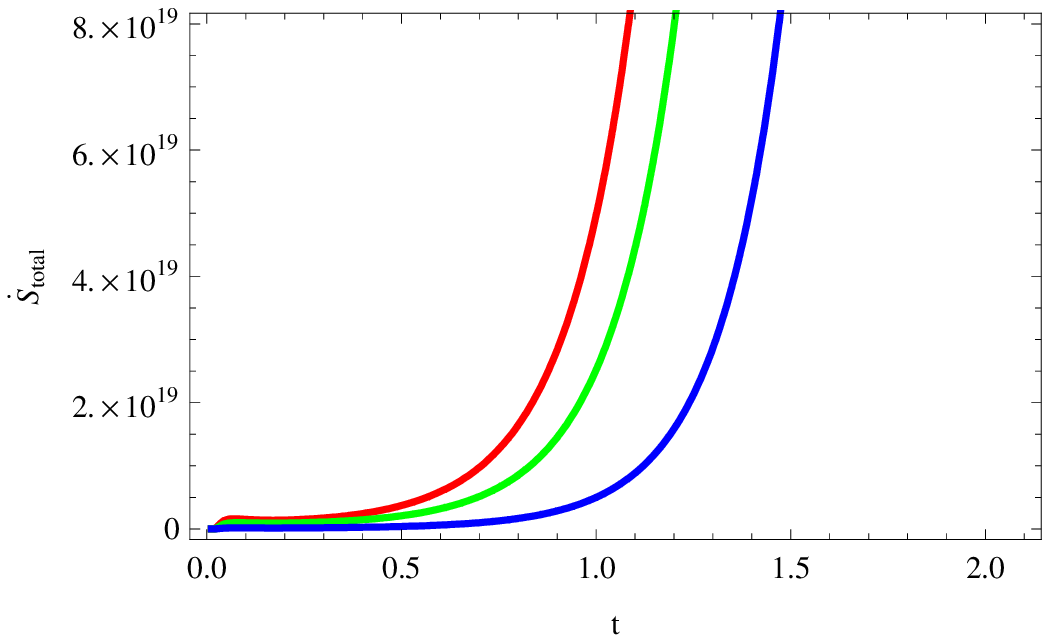}\\
\caption{ plots the time derivative of total entropy $\dot{S}_{total}$ against cosmic time $t$ for the universe enveloped by \emph{Event horizon} in the
$f(\cal R)$ gravity in the \emph{``Intermediate scenario"}. The red, green and blue lines correspond to $k=-1,~+1\text{and}~k=0$ respectively.}
\end{figure}
\begin{figure}
\includegraphics[height=2.0in]{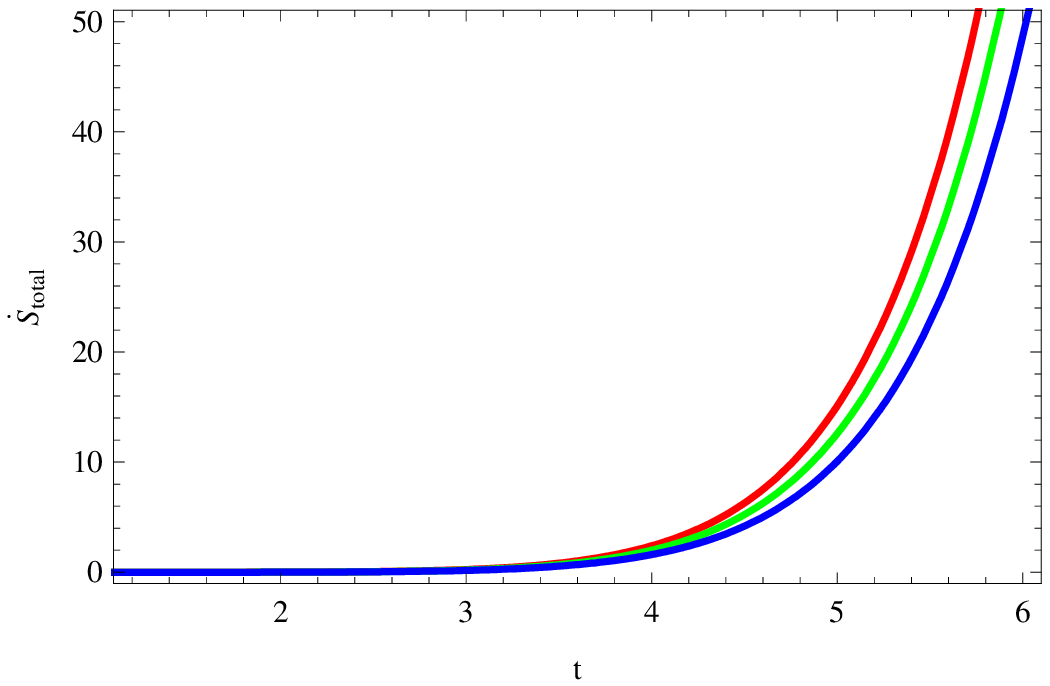}\\
\caption{ plots the time derivative of total entropy $\dot{S}_{total}$ against cosmic time $t$ for the universe enveloped by \emph{Event horizon} in the
$f(\cal R)$ gravity in the \emph{``Logamediate scenario"}. The red, green and blue lines correspond to $k=-1,~+1\text{and}~k=0$ respectively.}
\end{figure}

\section{Discussions}In the present work, we have investigated the
validity of generalized second law of thermodynamics in an universe enveloped by Hubble, apparent, particle and event horizons.Instead of considering FRW
universe governed by Einstein gravity we have considered a modified gravity in the form of $f(\cal R)$ gravity.We have chosen the scale factors in three
forms corresponding to emergent, intermediate and logamediate scenarios.While investigating the the validity of GSL of thermodynamics we have not taken
into account the first law of thermodynamics. The purpose being the investigation of the validity of GSL, we have computed the entropy on the horizon, as
well as, inside the horizon in the all twelve cases under consideration. We have kept the curvature of the universe under consideration.In all possible
three cases, we have examined the GSL for flat $(k=0)$, open $(k=-1)$ and closed $(k=1)$ universes. We have plotted the time derivative of the total
entropy $\dot{S}_{total}$ against the cosmic time $t$, in all of the cases under consideration.

In figure 1, 2 and 3 we have considered three choices of scale factors in an universe enveloped by hubble horizon and characterized by $f(\cal R)$ gravity.
In all of the three cases $\dot{S}_{total}$ is staying at positive level and exhibiting decaying behavior with passage of cosmic time $t$. This indicates
validity of GSL of thermodynamics in an universe characterized by $f(\cal R)$ gravity and enveloped by Hubble horizon. Moreover this holds irrespective of
the curvature of of the universe. It is further noted that for the intermediate and logamediate scenarios, the rate of decay of $\dot{S}_{total}$ is faster
than in the case of emergent scenarios. It is further noted that the decaying behavior is significantly influenced by the curvature of the universe in case
of logamediate scenario. Here From figure 3 we find that, in logamediate scenario, $\dot{S}_{total}$ falls very sharply in case of Flat universe $(k=0)$.
However this rate is much less in open $(k=-1)$ and closed $(k=1)$ universes.

In figure 4,5 and 6 we have considered apparent horizon. Although $\dot{S}_{total}$ stays at positive level in all these three cases, the nature of its
decay with cosmic time $t$ has varied with the choice of scale factor. It has been observed that, in the case of logamediate scenario (figure 6)
$\dot{S}_{total}$ has fallen very sharply irrespective of the curvature. Whereas, in the the case of emergent scenario (figure 4) the rate of change of
$\dot{S}_{total}$ is much lesser. In this case the decaying of $\dot{S}_{total}$ is very slow for flat (k=0). In the case intermediate scenario (figure 5),
the decaying of $\dot{S}_{total}$ is not significantly influenced by the curvature of the universe.

Figure 7,8 and 9 confirms the validity of GSL of thermodynamics in an universe characterized by $f(\cal R)$ gravity and enveloped by particle horizon. Here
$\dot{S}_{total}$ has not shown any significant dependence on the curvature of the universe. However, $\dot{S}_{total}$ behavior exhibited significant
changes in different scenarios of the universe. In the case of of emergent scenario it is increasing with cosmic time $t$, but in the case of intermediate
scenario in it decaying with cosmic time $t$. Although in the case of logamediate scenario (figure 9) $\dot{S}_{total}$ behaves differently from the other
scenarios. In figure 9 we can see that $\dot{S}_{total}$ is decaying with cosmic time $t$ after increasing upto a certain period of time.

In figures 10,11 and 12 we have plotted the time derivatives of total entropy for the universe in the emergent, intermediate and logamediate scenarios
respectively for the universe enveloped by the event horizon. These figures reveal that in $f(\cal R)$ gravity the GSL of thermodynamics is valid for all
the scenarios under consideration when we are assuming event horizon as the enveloping horizon of the universe.

Therefore, the rigorous study reported above reveals the validity GSL of thermodynamics in an universe govern by $f(\cal R)$ gravity. Irrespective of the
choice of scale factor, enveloping horizon and curvature of the universe, the time derivative of total entropy stays at the positive level. In the
reference \cite{bambageng}, the validity of GSL was investigated for $f(\cal R)$ gravity on the apparent horizon and it was shown that the GSL can be
satisfied in both phantom and non-phantom phases of the universe. The present study deviates from the said study in the respect for chosing a form of the
Ricci scalar and considering various forms of the scale factor available in the literature. Moreover, here we have not confined ourselves to the apparent
horizon only. We have also considered the other enveloping horizons like Hubble, particle and event horizons. In all of our cases under consideration, the
GSL of thermodynamics has been found to be satisfied.
\\\\
\section{Acknowledgements}
The first author wishes to thank the Inter-University Centre for Astronomy and Astrophysics (IUCAA), Pune, India for providing warm hospitality during a
scientific visit in January 2012, when part of the work was carried out. The second author sincerely acknowledges the Visiting Associateship provided by
IUCAA, Pune, India for the period of August 2011 to July 2014 to carry out research in General Relativity and Cosmology.
\\\\


\begin{thebibliography}{99}
\bibitem{sami} E. J. Copeland, M. Sami and S. Tsujikawa, \emph{Int. J. Mod. Phys. D}
\textbf{15} (2006) 1753.

\bibitem{Riess} A. G. Riess et al. (Supernova Search Team Collaboration),
Astron. J. 116, 1009 (1998).

\bibitem{Kluson} J. Kluson, \emph{Phys. Rev. D}
\textbf{81} (2010) 064028.

\bibitem{Nojirireview} S. Nojiri and S. D. Odintsov, \emph{Int. J. Geom. Meth. Mod. Phys.} \textbf{4} (2007)
115.

\bibitem{Mukhoyama} S. Mukhoyama, \emph{Class. Quantum  Grav.}, \textbf{27} (2010)
223101.

\bibitem{Paul} B. C. Paul., P.S. Debnath and S. Ghose, \emph{Phys. Rev. D} \textbf{79} (2009)
083534.


\bibitem{Nojiri1} S. Nojiri and S. D. Odintsov , \emph{Phys. Lett. B}
\textbf{657} (2007) 238.

\bibitem{Nojiri2} S. Nojiri and S. D. Odintsov , \emph{Phys. Rev.
D} \textbf{77} (2008) 026007.

\bibitem{chattopadhyay1} S. Chattopadhyay and U. Debnath,
\emph{International Journal of Modern Physics D} \textbf{20}
(2011) 1135.

\bibitem{Lima} J. A. S. Lima and J. S. Alcaniz, \emph{Phys. Lett. B} \textbf{600}
(2004) 191

\bibitem{Silva} R. Silva, J. S. Alcaniz and J. A. S. Lima ,\emph{International Journal of Modern Physics
D}\textbf{ 16} (2007) 469

\bibitem{Sheykhi} A. Sheykhi, Class. \emph{Quantum Grav.} \textbf{27 }(2010)
025007

\bibitem{chattopadhyay2} S. Chattopadhyay and U. Debnath,
\emph{International Journal of Modern Physics A} \textbf{30
}(2010) 5557

\bibitem{Setare2} M. R. Setare, \emph{Phys. Lett. B}\textbf{ 641 }(2006)
130

\bibitem{Setare3} M. R. Setare, \emph{J. Cosmol. Astropart. Phys.} \textbf{23}
(2007) 23.

\bibitem{Setare4} M. R. Setare and S.Shafei,\emph{J. Cosmol. Astropart.
Phys.} \textbf{9} (2006) 011

\bibitem{Elizalde} E. Elizalde, S. Nojiri, S. D. Odintsov and D.
Sáez-Gómez, \emph{The European Physical Journal C - Particles and
Fields } \textbf{70} (2010) 351.

\bibitem{Izquierdo} G. Izquierdo and D. Pavon, \emph{Phys.
Lett. B} \textbf{633} (2006) 420

\bibitem{jamil2} M. Jamil, E. N. Saridakis and M.R. Setare, \emph{JCAP}
\textbf{11}(2010) 032

\bibitem{Akbar2} M. Akbar and R-G. Cai, \emph{Phys. Lett. B}
\textbf{635}(2006) 7

\bibitem{Akbar3} M. Akbar and R-G. Cai, \emph{Phys. Lett. B}
\textbf{648}(2007) 243

\bibitem{Cai2} R-G. Cai and L-M. Cao, \emph{Phys. Rev. D} \textbf{75} (2007)
064008.

\bibitem{Wu} S-F. Wu, B. Wang and G-H. Yang, \emph{Phys. Lett. B}
\textbf{799} (2008) 330.

\bibitem{bambageng} K. Bamba and C-Q. Geng, \emph{Phys. Lett. B} \textbf{679}, 282 (2009)

\bibitem{Mukherjee}  S. Mukherjee, B. C. Paul, N. K. Dadhich,  S. D. Maharaj,  A. Beesham, \emph{Class. Quantum Grav.} \textbf{23} 6927 (2006)

\bibitem{barrow1} J. D. Barrow and N. J. Nunes, \emph{Phys. Rev. D} \textbf{76},043501 (2007)

\bibitem{barrow2} J. D. Barrow and  A. R. Liddle,  \emph{Phys. Rev. D} \textbf{47}, 5219 (1993)

\bibitem{ellis} G. F. R. Ellis, M. Madsen, \emph{Class. Quantum Grav.} \textbf{8}, 667
 (1991)

\bibitem{arundhati} A. Das, S. Chattopadhyay, U. Debnath, \emph{Foundations of Physics} DOI 10.1007/s10701-011-9600-1 (2011)

\bibitem{Nojiri3} S. Nojiri and S. D. Odintsov , \emph{Phys. Rev.
D} \textbf{68} (2003) 123512.

\end{thebibliography}
\end{document}